\renewcommand{\a}{\alpha}
\renewcommand{\b}{\beta}

\renewcommand{\d}{\delta}
\newcommand{\D}{\Delta}

\newcommand{\ep}{\epsilon}

\newcommand{\La}{\Lambda}

\newcommand{\tr}{\rm tr}

\newcommand{\p}{\partial}

\newcommand{\non}{\nonumber}

\newcommand{\lb}{\left(}
\newcommand{\rb}{\right)}
\renewcommand{\le}{\left\lbrack}
\newcommand{\re}{\right\rbrack}

\newcommand{\beq}{\begin{equation}}
\newcommand{\eeq}{\end{equation}}
\newcommand{\beqa}{\begin{eqnarray}}
\newcommand{\eeqa}{\end{eqnarray}}

\newcommand{\prd}[1]{{ \it Phys.~Rev.}~{\bf D{#1}}}
\newcommand{\prl}[1]{{ \it Phys.~Rev.~Lett.}~{\bf {#1}}}
\newcommand{\plb}[1]{{ \it Phys.~Lett.}~{\bf {#1B}}}
\newcommand{\pla}[1]{{ \it Phys.~Lett.}~{\bf {#1A}}}
\newcommand{\npb}[1]{{ \it Nucl.~Phys.}~{\bf B{#1}}}

\newcommand{\rmp}[1]{{ \it Rev.~Mod.~Phys.}~{\bf {#1}}}
\newcommand{\cmp}[1]{{\it Comm.~Math.~Phys}~{\bf {#1}}}

\newcommand{\vn}{\vec n}
\newcommand{\vt}{\vec t}
\newcommand{\vX}{\vec X}

\documentstyle[12pt]{article}
\begin{document}
\baselineskip 18pt

\begin{titlepage}
\renewcommand{\thefootnote}{\fnsymbol{footnote}}

\begin{flushright}
\parbox{1.in}
{
 IFT-18/93\\
 September 1993}
\end{flushright}
\vspace*{.5in}
\begin{centering}
{\Large  AN EFFECTIVE 3D RIGID STRING AT $\theta=\pi$\footnote{
Work supported, in part,
by Polish Government Research Grant KBN 
}}\\
\vspace{2cm}
{\large        Jacek Pawe\l czyk}\\
\vspace{.5cm}
        {\sl Institute of Theoretical Physics, Warsaw  University,\\
        Ho\.{z}a 69, PL-00-681 Warsaw, Poland.}\\
\vspace{.5in}
\end{centering}
\begin{abstract}
An effective sigma model describing behavior of the 3d rigid string
with a $\theta$-term at $\theta=\pi$ is proposed. It contains
non-perturbative corrections resulting from summation over different
genera of the 2d surfaces. The effective theory is the $SU(2)$ WZW model
coupled to the Nambu-Goto action. RG analysis shows the existence of
a IR fixed point at  which the normal to the surface has long range
correlations. A similar model can describe critical behaviour of the
3d Y-M fields or the Ising model.
\end{abstract}

\end{titlepage}

\setcounter{footnote}{0}
\renewcommand{\thefootnote}{\arabic{footnote}}

It is strongly believed that the dynamics of gauge fields can be described
in terms of a string theory.
The $1/N_c$  expansion \cite{thooft}, the lattice strong
coupling expansion \cite{lat} and very recently the arguments by
D.Gross concerning 2d QCD \cite{gross} strongly support this idea.
String theory (random
surfaces theory) is also believed to describe the critical behavior of
the 3d Ising model \cite{polising,polbook}.
In the dual picture the 3d Ising model is
the 3d $Z_2$ gauge theory \cite{savit}.
Unfortunately no satisfactory string picture was constructed till now
for any of these gauge theories \cite{poldot,polrig}.
In \cite{poldot} it was  claimed that
it could be a kind of NSR fermionic string, but the obtained
discrete version was not suitable for the continuum limit.

It is known that a gas of
self-avoiding surfaces in a special 3d lattice is in the same
universality class as the Ising model \cite{david1}.
Recent lattice simulations \cite{gliozzi}
show the  necessity of the non-perturbative renormalization.
 The surfaces at small scales look like sponge
due to large number of microscopic handles. Yet the summation over
all genera "smoothes out" the surface on large scales.
This indicates the appearance of the non-perturbative renormalization
effects due to the summation over genera. One can expect that
the renormalization may have drastic impact on the low energy action
describing the dynamics  of the surface.

Similar ideas appeared in theoretical works.
J.Distler \cite{distler} analyzed the Ising model again arguing that
the formula for the random surface partition function could be
\beq
Z[\mu]=\sum_{\{M\}}(-1)^{w_2(M)}e^{-\mu A[M]}
\label{dis}
\eeq
where $\{M\}$ is a sum over all immersions of 2d surface into 3d space
$R^3$, $A[M]$ is the area of the manifold $M$ and
$w_2$ is the second Stifel-Whitney class which for the closed
surfaces is the modulo two reduction  of the Euler characteristic:
$w_2(M)=\chi(M)$ mod 2.
The claim  is that the alternating factor introduces
cancellation between topologically different, but physically equivalent
random surfaces.  Such a factor may drastically change the low energy
behavior of (\ref{dis}) due to enormous cancellation between various
topological sectors of the model. This kind of non-perturbative
renormalization takes place in the 2d Ising model
whose critical behavior is described by the fermionic random walk.
After integrating out the world-line fermions one obtains a random walk with
an alternating
factor $(-1)^n$ where $n$ is the self-intersection number of the walk.
World-sheet fermions may play the same
role \cite{wiegmann}. Moreover they also produce
dependence on the extrinsic curvature of the immersed surface. Thus it
is quite possible that the correct action for the 2d model should
include the extrinsic curvature.

In this paper we consider a model which incorporates all the
mentioned above features.
Thus the proposed (short distance) action is
\beq
S[X]=\int d^2x\,\sqrt{g} \le\mu +\frac{1}{2\alpha}
g^{ab}(\p_a \vn)(\p_b \vn) +\frac{i}{8}R\re,
\label{sigma}
\eeq
where $\vn^2=1$ and in addition
\beq
\p_a\vX\,\vn=0.
\label{constr}
\eeq
i.e. $\vn$ is normal to the immersed surface.
The metric appearing above is the  induced metric $g_{ab}\equiv
\p_a\vX\p_b\vX$.
The second term in the formula above is the extrinsic curvature term
(or rigidity) \cite{polrig,pfk}
which up to curvature $R$ is $(\D\vX )(\D \vX )$.
Here we can note the importance  of the constraints  (\ref{constr}),
which make (\ref{sigma}) ''naively'' renormalizable.\footnote{
The equivalence of (\ref{sigma}) and the rigid string was checked
perturbatively up to one-loop.} A similar model with
dynamical metric was proposed in  \cite{polles}.

The last term of the formula (\ref{sigma}) is proportional to
the Euler characteristic. It is
the source of the alternating factor analogous  to that of (\ref{dis})
if one takes  into account only the orientable surfaces. This
contrasts with (\ref{dis}) where the sum runs also over the nonorientable
surfaces.

Let us note that we can express (up to sign) the Euler
characteristic in terms of the normal field:
\beq
\chi(M)=\frac{1}{4\pi}\int_M d^2x \ep^{ab}\vn\,(\p_a\vn\times \p_b\vn)
\label{euler}
\eeq
This is just twice the winding number of the map $S^2\to S^2$.
Thus the last two terms of (\ref{sigma})  can be considered as
the $O(3)$ sigma model with the $\Theta$ term at $\Theta=\pi$.
In this way the topologically different sectors of the $O(3)$ model
correspond to Riemann surfaces of even Euler characteristic.

Without the $\Theta$ term  the $O(3)$ sigma model is massive \cite{bre}.
For  $\Theta=\pi$  it is claimed to have non trivial IR fixed
point at which the theory is described by the
$k=1$ $SU(2)$ WZW model \cite{witten}. This remarkable fact was conjectured
(Haldane conjecture) by many authors \cite{haldane,affleck,zamolodchikov}.
The non-perturbative
contributions from different topological sectors play crucial role
in  this change of behaviour of the $O(3)$ model. One may expect that
a similar mechanism would work in the case of the model (\ref{sigma}).
Of course here the situation is much more complicated because the
constraints (\ref{constr}) give rise to non trivial coupling between the
immersions $\vX$ and the normal field $\vn$.

Hereafter we assume that the non-perturbative effects
change the $O(3)$ sigma model part of (\ref{sigma}) to
the $SU(2)$ WZW model or its modifications. The source of modifications
are the constraints (\ref{constr}).
Now we face the problem: what is the proper expression for the normal vector
in terms of $O(3)\propto SU(2)$ group element ?
The line of reasoning given in \cite{affleck} starts from the Hubbard
model \cite{fradkin} which describes the dynamics of electrons in
(1+1) dimensions.
Its nonabelian bosonization leads to the $k=1$ $SU(2)$ WZW model.
In this construction the vector $\vn$ is part of the spin
of the electron ${\vec S}$ which in bosonization language has
the low energy limit
${\vec S}\propto \tr(h{\vec \tau})$. Thus it is seems natural to
identify $\vn$ with the r.h.s. of the last formula.
\beq
{\vec n}\propto \tr(h{\vec \tau})
\label{vecn}
\eeq

The model described above is strongly interacting. Its solution may be
not an easy task.
In what follows we are going  to perform perturbative calculations
taking the large $k$. From this point of view (\ref{vecn})
leads to troublesome, from perturbative point of view,
interactions.
One may expect it because the above reasoning invoke k=1 WZW model
which is strongly interacting.
Thus instead of (\ref{vecn}) we shall use a proper representation
in the $k\to \infty$ limit:
\beq
\vn=\frac{1}{2}\tr({\vec \tau}\, h \tau^3 h^{-1}),
\label{normal}
\eeq
with $h$ being $SU(2)$ group element. Its conformal dimension is
$2/(k+2)$ thus equals zero for $k \to \infty$. With (\ref{normal})
the rotation group is $SU_L(2)$ , the correct group in this limit.

In this paper we shall discuss the simplest low energy theory.
The only obvious symmetry we want  to  preserve is the global $O(3)$ of
simultaneous rotations of the vectors ($\vn,\p_a\vX$).
The choice is not
rich because the constraint (\ref{constr}) indicates the following
identity:
\beq
n^\mu n^\nu=\d^{\mu\nu}-g^{ab}\p_a\vX^\mu\p_b\vX^\nu
\eeq
Together with (\ref{constr}) this eliminates  powers of the normal
field. We also note that the normal vector (\ref{normal})
does not change under the $U_R(1)$ subgroup of the $SU(2)$ thus the
constraints (\ref{constr}) break the symmetry of the sigma model.
The
final form of the model we shall consider is
\beqa
S[X]&=&\int_{S^2} d^2x \le\mu \sqrt{g}-\sqrt{g}g^{ab}\lb \frac{1}{\alpha}
\tr[(h^{-1}D_a h)(h^{-1}D_b h)]+\frac{1}{2r}\tr(j_a\tau^3)tr(j_b\tau^3)\rb\re
\non\\&+&\frac{ik}{12\pi}\int_B tr(j\wedge j\wedge j)
\label{model}
\eeqa
where $h$ is $SU(2)$ element.
$$
j_a=h^{-1}\p_a h,\quad D_a h=\p_a h+A_a h\tau^3, \quad
A_a=-\frac{1}{2}\tr (j_a\tau^3)
$$
 together with the nontrivial constraint (\ref{constr}).

In the following we shall
consider RG behavior of the coupling constants of (\ref{model}).
The model  is a  modification of  the rigid string \cite{polrig,pfk}.
The rigid string has RG behaviour which shows  irrelevance of the
rigidity for large
 distances at least for small couplings $\a$. If  so the low
energy theory describing the dynamics of string is the Nambu-Goto
action with its well know diseases. The rigidity is irrelevant
mainly because the $O(3)$ model  is massive.
The forthcoming analysis
shows  quite opposite behaviour of (\ref{model}):
the coupling $\a$ has an IR fixed point
signaling the long range correlations of normals. The problem is also
interesting on its own as an
example of coupling of a CFT to the induced metric and the
Nambu-Goto action (see also (\cite{kavalov}) .

We are going to use the background field method.
Before one proceeds with calculation the gauge freedom (2d
reparameterization invariance) has to be fixed.  We choose to work in
so-called normal gauge \cite{david} in which the quantum part of the
$\vX$ field has only the normal component i.e.\footnote{In order not
to proliferate  notations all the quantities (except the quantum
fields $\pi,\;\xi$) on the r.h.s. of the eqs (\ref{xback},\ref{pi})
 will be  background quantities.}
\beq
\vX\to \vX+\xi\vn,\quad h\to h\, e^{i\pi}
\label{xback}
\eeq
where $\xi,\,\pi\equiv {\vec \pi}{\vec \tau}$  are the quantum fields.
In order to do perturbative
calculation we need to solve the constraints first.
Up to terms quadratic in quantum fields
\beq
\pi^a=-\frac{1}{2}(E^{ab}\p_b\xi+E^{ab}K_b^c\xi\p_c\xi-g^{ab}\pi^3\p_b\xi)
\label{pi}
\eeq
where\footnote{
We change indices from the flat ($\a,\b,...$) to curved ($a,b,...$)
with help of the induced background "zweibien" $e_{a\a}\equiv \p_a\vX
\vt_\a$  and its inverse $e^a_{\;\a}$. The vectors $\vt_\a$ ($\a=1,2$)
are tangent to the surface. They can be expressed by the $SU(2)$
elements as $\vt_\a=\frac{1}{2}\tr ({\vec \tau} h \tau^\a h^{-1})$.}
$E^{ab}\equiv \ep^{\a\b}e^a_\a e^b_\b$.
In fact the last two terms of (\ref{pi}) do not contribute:
the first contains too few
derivatives of the quantum fields , the second is irrelevant
because there is no mixing between $\xi$ and $\pi^3$.

Now one can expand the action (\ref{model}) in
powers of quantum fields and calculate the one-loop renormalization of
the couplings of the model. This kind of calculations have been done
many times thus we refer the reader to the literature
\cite{pfk,pelti,david}. We just state the final result and discuss it.
\beq
\frac{2}{\a(\mu)}= \frac{2}{\a(\La)}+3\lb \lb \frac{k}{8\pi}
\rb^2\a r-1+\frac{\a}{4r}\rb \log\frac{\La}{\mu}
\label{ren}
\eeq
{}From the above one directly infers that the flow of the coupling $\a$ has
an IR stable point given  by
\beq
\a^*=\frac{1}{\lb \frac{k}{8\pi}\rb^2 r+\frac{1}{4r}}
\label{fix}
\eeq
For $\a^*=r$ this gives $(k\a^*/8\pi)^2=3/4$ what is very close the
pure WZW result \cite{witten}. We expect that $k$ is close to 1,
thus for reasonable $r$,  the coupling $\a^*$ is a big number
far from the region of applicability of the perturbative calculations.
It does not mean
that we should not believe in existence of the IR fixed point
(although its value may be completely different) because the very reason
for its appearance is the WZ term just as it happens in the WZW model.

It is interesting to note that the couplings $k$ and $r$ are not
renormalized. The reason  for such a behavior of $k$ is, as usual,
its topological origin.
The technical  reason  for non-renormalization of $r$ (at one-loop)
is group theoretical. The current $j^3_a$ is an isospin zero
member of $SU(2)$ triplet thus it couples only to
the appropriate combination of the $\pi$ fields (eq.(\ref{xback})) namely
$j^3_a\ep^{\a\b}\p_a\pi^\a \pi^\b$.
{}From eq.(\ref{xback}) the $\pi$ fields  are proportional to
derivatives of $\xi$. This in turn makes one-loop
contributions zero due to the epsilons under integrals.

The string tension is also renormalized. At one-loop the
WZ term  does not contribute thus the renormalization is similar to
what one usually gets for the rigid string. Thus the RG flow in the
($\a,\,\mu$) plane looks the same as that obtained in \cite{pelti,david}.
Moreover the detailed discussion of the theory at the fixed point is the same.
We refer the reader  to these publications.
What we want to emphasize here is that at the critical point the
surface is smooth. There are long range correlations between normals to
the surface. One may expect that the tachyon problem disappears
in such a theory.
\vskip.5cm

{\it Conclusions and comments.}
Our analysis left a lot of questions to be answered.
The basic problem is the analysis of the model for $k=1$.
Without this one can not be sure if the action (\ref{model})
has been chosen properly. It could be useful
to have more direct  arguments for it.
The result (\ref{fix}) is reliable only for large
enough $k$, although it may happen that it is close to the exact one
even for small $k$ as it is for the WZW models.
This could be checked by the next-to-leading contributions. Despite this the
critical indices of the low $k$ models are expected to be completely different
then those of the large $k$ models.
The next group of problems lies in the use of the induced metric.
Maybe, one should use the Polyakov action instead the Nambu-Goto and
work with dynamical 2d  gravity \cite{polles}?
These problems are the subject of current research.
\vskip.5cm
{\bf Acknowledgment}

I want to thank  R.Budzy\'nski, M.Spali\'nski
for their kind interest in this work and  A.Kavalov for presentation of
his results.

\end{document}